\documentclass[10pt,aps,prd,twocolumn,groupedaddress,floatfix,a4paper]{revtex4}
\usepackage{graphicx,amsmath,amssymb}
\usepackage{dcolumn}
\usepackage{bm}
\usepackage{longtable}
\usepackage{color}
\usepackage{float}
\begin{document}

\title{Was the cosmic ray burst detected by the GRAPES-3 on 22 June 2015 caused by transient weakening of geomagnetic field or by an interplanetary anisotropy?}

\author{P.K.~Mohanty}
\affiliation{Tata Institute of Fundamental Research, Homi Bhabha Road, 
Mumbai 400005, India}
\altaffiliation{The GRAPES-3 Experiment, Cosmic Ray Laboratory, Raj Bhavan, Ooty 643001, India}

\author{K.P.~Arunbabu}
\affiliation{Tata Institute of Fundamental Research, Homi Bhabha Road,
Mumbai 400005, India}
\altaffiliation{The GRAPES-3 Experiment, Cosmic Ray Laboratory, Raj Bhavan, Ooty 643001, India}

\author{T.~Aziz}
\affiliation{Tata Institute of Fundamental Research, Homi Bhabha Road,
Mumbai 400005, India}
\altaffiliation{The GRAPES-3 Experiment, Cosmic Ray Laboratory, Raj Bhavan, Ooty 643001, India}

\author{S.R.~Dugad}
\affiliation{Tata Institute of Fundamental Research, Homi Bhabha Road,
Mumbai 400005, India}
\altaffiliation{The GRAPES-3 Experiment, Cosmic Ray Laboratory, Raj Bhavan, Ooty 643001, India}

\author{S.K.~Gupta}
\email[]{gupta.crl@gmail.com}
\affiliation{Tata Institute of Fundamental Research, Homi Bhabha Road,
Mumbai 400005, India}
\altaffiliation{The GRAPES-3 Experiment, Cosmic Ray Laboratory, Raj Bhavan, Ooty 643001, India}

\author{B.~Hariharan}
\affiliation{Tata Institute of Fundamental Research, Homi Bhabha Road,
Mumbai 400005, India}
\altaffiliation{The GRAPES-3 Experiment, Cosmic Ray Laboratory, Raj Bhavan, Ooty 643001, India}

\author{P.~Jagadeesan}
\affiliation{Tata Institute of Fundamental Research, Homi Bhabha Road,
Mumbai 400005, India}
\altaffiliation{The GRAPES-3 Experiment, Cosmic Ray Laboratory, Raj Bhavan, Ooty 643001, India}

\author{A.~Jain}
\affiliation{Tata Institute of Fundamental Research, Homi Bhabha Road,
Mumbai 400005, India}
\altaffiliation{The GRAPES-3 Experiment, Cosmic Ray Laboratory, Raj Bhavan, Ooty 643001, India}

\author{S.D.~Morris}
\affiliation{Tata Institute of Fundamental Research, Homi Bhabha Road,
Mumbai 400005, India}
\altaffiliation{The GRAPES-3 Experiment, Cosmic Ray Laboratory, Raj Bhavan, Ooty 643001, India}

\author{P.K.~Nayak}
\affiliation{Tata Institute of Fundamental Research, Homi Bhabha Road,
Mumbai 400005, India}
\altaffiliation{The GRAPES-3 Experiment, Cosmic Ray Laboratory, Raj Bhavan, Ooty 643001, India}

\author{P.S.~Rakshe}
\affiliation{Tata Institute of Fundamental Research, Homi Bhabha Road,
Mumbai 400005, India}
\altaffiliation{The GRAPES-3 Experiment, Cosmic Ray Laboratory, Raj Bhavan, Ooty 643001, India}

\author{K.~Ramesh}
\affiliation{Tata Institute of Fundamental Research, Homi Bhabha Road,
Mumbai 400005, India}
\altaffiliation{The GRAPES-3 Experiment, Cosmic Ray Laboratory, Raj Bhavan, Ooty 643001, India}

\author{B.S.~Rao}
\affiliation{Tata Institute of Fundamental Research, Homi Bhabha Road,
Mumbai 400005, India}
\altaffiliation{The GRAPES-3 Experiment, Cosmic Ray Laboratory, Raj Bhavan, Ooty 643001, India}

\author{M.~Zuberi}
\affiliation{Tata Institute of Fundamental Research, Homi Bhabha Road,
Mumbai 400005, India}
\altaffiliation{The GRAPES-3 Experiment, Cosmic Ray Laboratory, Raj Bhavan, Ooty 643001, India}

\author{Y.~Hayashi}
\affiliation{Graduate School of Science, Osaka City University, 558-8585 Osaka, Japan}
\altaffiliation{The GRAPES-3 Experiment, Cosmic Ray Laboratory, Raj Bhavan, Ooty 643001, India}

\author{S.~Kawakami}
\affiliation{Graduate School of Science, Osaka City University, 558-8585 Osaka, Japan}
\altaffiliation{The GRAPES-3 Experiment, Cosmic Ray Laboratory, Raj Bhavan, Ooty 643001, India}

\author{P.~Subramanian}
\affiliation{Indian Institute of Science Education and Research, Pune 411021, India}
\altaffiliation{The GRAPES-3 Experiment, Cosmic Ray Laboratory, Raj Bhavan, Ooty 643001, India}

\author{S.~Raha}
\affiliation{Bose Institute, 93/1, A.P.C. Road, Kolkata 700009, India}
\altaffiliation{The GRAPES-3 Experiment, Cosmic Ray Laboratory, Raj Bhavan, Ooty 643001, India}

\author{S.~Ahmad}
\affiliation{Aligarh Muslim University, Aligarh 202002, India}
\altaffiliation{The GRAPES-3 Experiment, Cosmic Ray Laboratory, Raj Bhavan, Ooty 643001, India}

\author{A.~Oshima}
\affiliation{College of Engineering, Chubu University, Kasugai, Aichi 487-8501, Japan}
\altaffiliation{The GRAPES-3 Experiment, Cosmic Ray Laboratory, Raj Bhavan, Ooty 643001, India}

\author{S.~Shibata}
\affiliation{College of Engineering, Chubu University, Kasugai, Aichi 487-8501, Japan}
\altaffiliation{The GRAPES-3 Experiment, Cosmic Ray Laboratory, Raj Bhavan, Ooty 643001, India}

\author{H.~Kojima}
\affiliation{Faculty of Engineering, Aichi Institute of Technology, Toyota City, Japan}
\altaffiliation{The GRAPES-3 Experiment, Cosmic Ray Laboratory, Raj Bhavan, Ooty 643001, India}

\begin{abstract}
The GRAPES-3 muon telescope in Ooty, India had claimed
detection of a 2\,hour\,(h) high-energy ($\sim$20\,GeV)
burst of galactic cosmic-rays (GCRs) through a $>$50$\sigma$
surge in GeV muons, was caused by reconnection of the
interplanetary magnetic field (IMF) in the magnetosphere
that led to transient weakening of Earth's magnetic shield.
This burst had occurred during a G4-class geomagnetic storm
(storm) with a delay of $\frac{1}{2}$h relative to the coronal
mass ejection (CME) of 22 June 2015 \cite{Mohanty16}. However,
recently a group interpreted the occurrence of the same burst
in a subset of 31 neutron monitors (NMs) to have been
the result of an anisotropy in interplanetary space
\cite{Evenson17} in contrast to the claim in
\cite{Mohanty16}. A new analysis of the GRAPES-3 data with a
fine 10.6$^{\circ}$ angular segmentation shows the speculation
of interplanetary anisotropy to be incorrect, and offers
a possible explanation of the NM observations. The observed
28\,minutes (min) delay of the burst relative to the CME can
be explained by the movement of the reconnection front from
the bow shock to the surface of Earth at an average speed
of 35\,km/s, much lower than the CME speed of 700\,km/s. This
measurement may provide a more accurate estimate of the start
of the storm.
\end{abstract}

\maketitle

\section{Introduction}\label{intro}
\vskip -0.25in
It is well-known that the geomagnetic field (GMF) \cite{Olsen15}
acts as the first line of defense by shielding the Earth from
energetic galactic cosmic rays (GCRs) through
magnetic deflection out to several Earth radii (R$\rm_{E}$)
\cite{Dorman09}. The CMEs produced by the Sun are large bodies
of plasma containing highly turbulent magnetic fields that are
driven into heliosphere from the solar corona \cite{Webb12}.
The interaction of this turbulent magnetized plasma
with the GCRs produces a modulation of GCRs that offers
an excellent probe of the space weather \cite{Kudela00}.
The CMEs being a major driver of the space weather can have a
large societal impact by triggering severe storms with potential
to disrupt the space-, and ground-based communications.
The largest storm in the recorded history was observed by
Carrington in 1859 that disrupted the old rugged communication
system of telegraph lines for several hours
\cite{Carrington59,Liu14}. But the occurrence of a similar event
today would surely cripple the modern infrastructure of mobile
phones, computer networks on the ground, and the satellites in
space. This is primarily due to an ever increasing miniaturization
of the present day electronic devices that are unlikely to survive
the high radiation environment created by a Carrington-class storm
\cite{NASA08,OECD11}.

Reduction of the GCR intensity (GCRI) lasting several days
due to turbulent IMF in a CME known as a Forbush decrease
(FD), have been observed for decades \cite{Forbush38}. The
episodes of short term increase ($\sim$h) in the GCRI due to
the lowering of the geomagnetic `cutoff rigidity R$\rm _c$'
\cite{Smart05} were also reported earlier
\cite{Kondo62,Kudo87,Dorman09}. The GRAPES-3 muon telescope had
reported the detection of a 2\,h burst starting 22 June 2015
19:00 UT, that was strongly correlated (94\%) with a 40\,nT
surge in the IMF. A unique feature of this burst was a delay
of about $\frac{1}{2}$h relative to the IMF surge for which
no satisfactory explanation was offered earlier. Monte Carlo
simulations to reproduce this burst required compression of
the IMF to 680\,nT spread over several times the volume of Earth,
followed by reconnection with the GMF leading to lower R$\rm _c$
had generated this burst \cite{Mohanty16}.

Recently, a group after examining the data from 31 NMs
spread across the world had reported that only ten of them
displayed a sharp feature similar to the GRAPES-3 burst. They
claimed that any variation in R$\rm _c$ caused by
the IMF will be a global process, which is inconsistent
with the detection of the burst by only a subset of 31 NMs.
They had offered an alternative interpretation of the
GRAPES-3 burst as being a manifestation of an interplanetary
anisotropy (IA) \cite{Evenson17}. In the present work this
interpretation is critically examined through a new analysis
of the GRAPES-3 data with a finer angular segmentation of
10.6$^{\circ}$. 

\begin{figure}[t]
\begin{center}
\includegraphics*[width=0.48\textwidth,angle=0,clip]{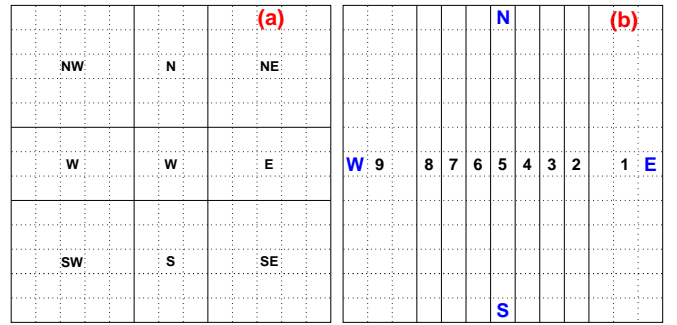}
\vskip -0.05in
\caption{\label{fig1} (a) 9 directions (FOV\,=\,2.3\,sr.) used
                      earlier \cite{Mohanty16}. (b) a new
                      combination of 169 directions labeled 1
                      through 9 with an angular segmentation
                      of 10.6$^{\circ}$ used here.}
\end{center}
\end{figure}

The large area (560\,m$^2$) GRAPES-3 muon telescope,
hereafter called, `telescope' is located in Ooty,
India. The telescope experiences high cutoff rigidities
(15--27\,GV) due to its near-equatorial location
(11.4$^{\circ}$N). The telescope measures the intensity of
$\geq$1\,GeV atmospheric muons produced by the GCRs, along
13$\times$13\,=\,169 directions in the sky
\cite{Hayashi05,Nonaka06}. Thus, the muons serve as a
good GCR proxy, and therefore, the terms `muon
intensity', and GCRI will be used interchangeably. In our
previous work these 169 directions were combined into nine
directions, covering a field of view (FOV) of 2.3\,sr as
shown in Fig.\,\ref{fig1}a \cite{Mohanty16}. However, here
the 169 directions are combined in a different configuration,
labeled 1 through 9 from East to West with a mean angular
segmentation of (10.6$^{\circ}$\,$\pm$\,1.1$^{\circ}$) as shown
in Fig.\ref{fig1}b. This scheme was used because an IA will
first appear in direction 1, and then progressively later in
directions 2, through 9. A maximum time delay of 5.5\,h is
expected to occur between directions 1, and 9.

As before, in the present analysis also the CME
parameters including the solar wind speed (V$\rm _{SW}$), the
IMF components B$\rm _x$, B$\rm _y$, B$\rm _z$ measured by
the WIND spacecraft located at L1 (1.5$\times$10$
^6$\,km from Earth) from OMNIWeb were used \cite{Omniweb}.
The WIND data from OMNIWeb were already shifted in time to the
bow shock nose to account for the propagation delay from
the spacecraft \cite{Bargatze05}.

\section{Detection of the GRAPES-3 burst by other experiments}\label{com}
\begin{figure*}[t]
\begin{center}
\includegraphics*[width=0.95\textwidth,angle=0,clip]{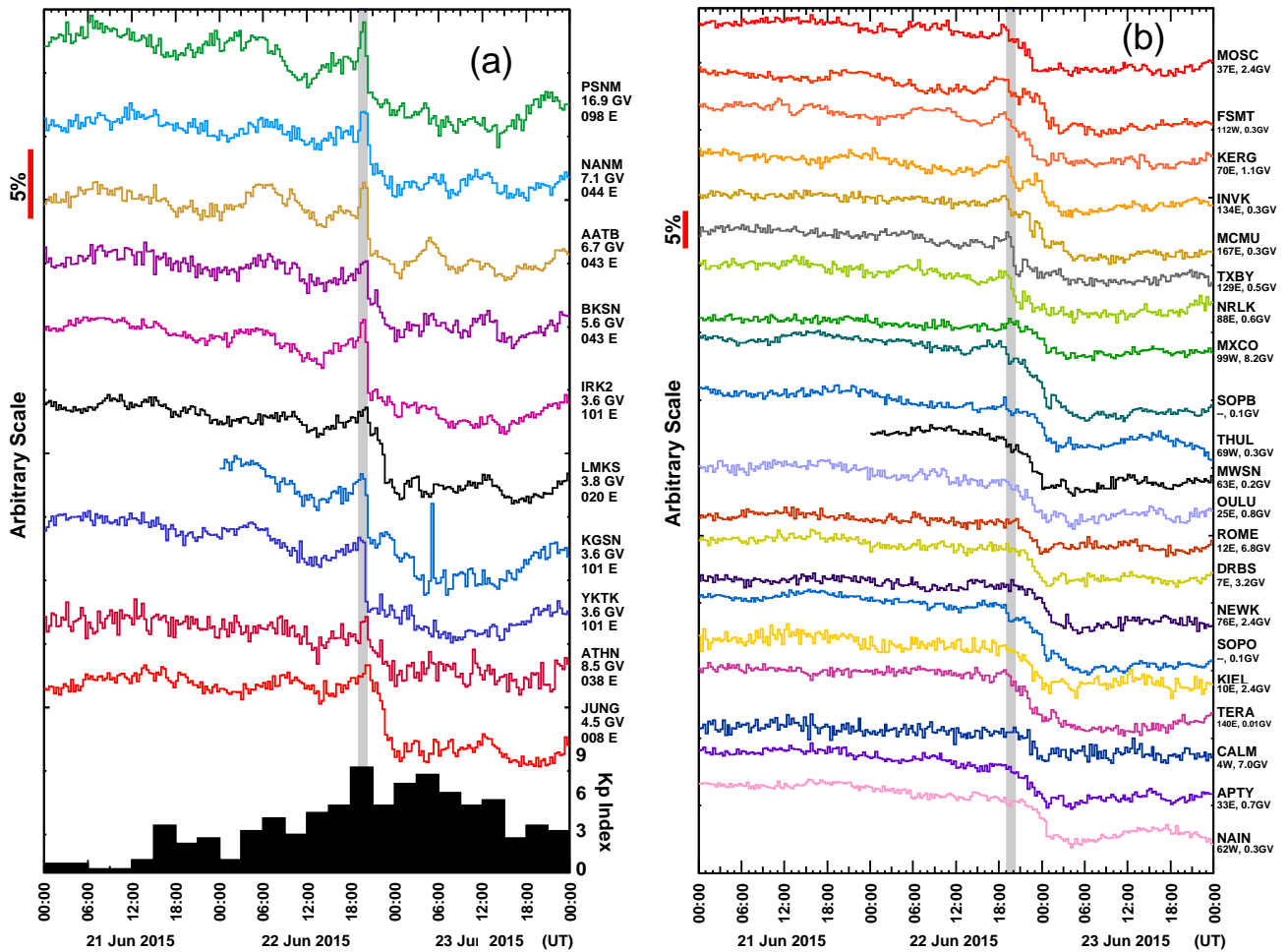}
\vskip -0.05in
\caption{\label{fig2} (a) Response of ten NMs with largest increase 
                      coincident with GRAPES-3 burst on 22 June
                      2015. Kp index of geomagnetic activity shown
                      at bottom, (b) Response of 21 NMs with no
                      coincident activity. Duration of GRAPES-3
                      burst indicated by `Gray' shading.}
\end{center}
\vskip -0.3in
\end{figure*}

\vskip -0.2in
Recently, a group had reported detection of the burst
from ten NMs as shown in Fig.\,\ref{fig2}a, while the remaining
21 NMs did not show any burst-like activity as seen from
Fig.\,\ref{fig2}b, which is a exact reproduction of
Fig.\,2 from \cite{Evenson17}. One feature that stands out is
the fact that the mean cutoff rigidity of the ten NMs
detecting the burst was a high 6.4\,GV, compared to 1.8\,GV for
the remaining 21 NMs that did not record the burst.
This important feature will be discussed in some detail
at the end of this section.

\begin{figure}[]
\begin{center}
\includegraphics*[width=0.48\textwidth,angle=0,clip]{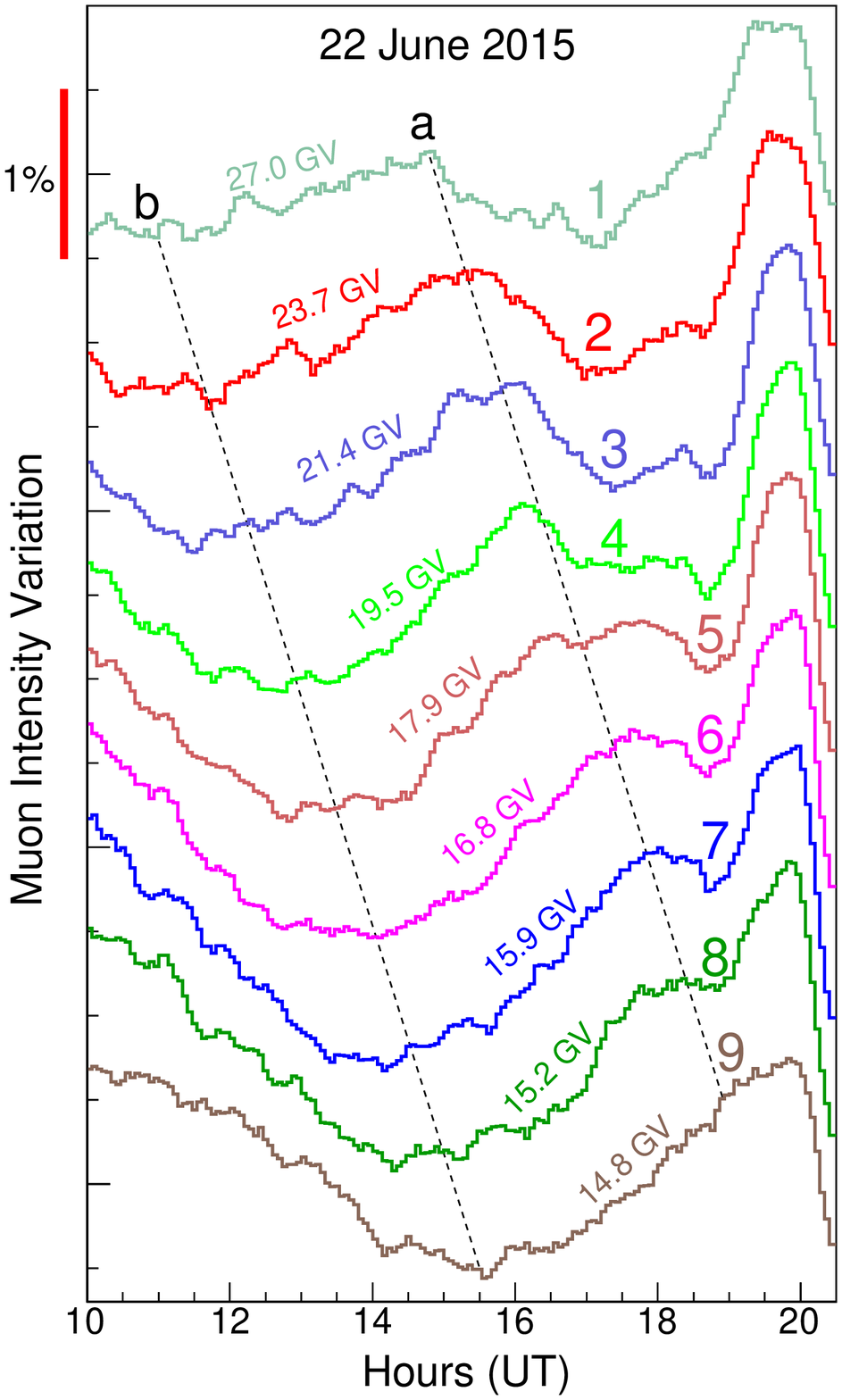}
\vskip -0.05in
\caption{\label{fig3} Muon intensity variation in 9 directions
                      observed by GRAPES-3 on 22 June 2015.
                      Progressive shift of anisotropy `peak'
                      is seen going from East to West (labeled
                      1 through 9). The shift in location of
                      anisotropy `peaks' and `valleys' are
                      marked by inclined dashed lines labeled
                      `a', and `b', respectively.}
\end{center}
\vskip -0.3in
\end{figure}

The muon intensity corrected for the instrumental, and
atmospheric pressure variations \cite{Mohanty15} contains
the modulation due to the FD, and the IA. For
studying the IA, the muon intensity was smoothed by taking
a 16\,min running average. The resultant intensities are
shown in Fig.\,\ref{fig3} for the nine directions labeled
1 through 9, progressing from East to West as shown in
Fig.\,\ref{fig1}b. The muon intensity shows a shift of the
peak to later times for directions 2 through 9 just as
expected for an IA. The locations of the IA peaks in
Fig.\,\ref{fig3} are indicated by an inclined line labeled
`a' and the valleys by `b'. The R$\rm _c$ for these directions
varies from a high value of 27.0\,GV for 1 to a low of
14.8\,GV for 9. Interestingly, this decrease in R$\rm _c$
is associated with a steady increase of the IA amplitude from
1 to 9. The 2\,h burst is clearly visible in each of the
nine directions, and peaks at 20\,UT in all cases in contrast
to the IA that peaks at progressively later times for
the western directions, eventually merging with the burst
in direction 9.

The IA amplitude shows a clear power-law dependence on
R$\rm _c$ with a spectral index of -1.5 for the rigidity
range 15--27\,GV. If such a dependence were to continue to
lower rigidities, the IA could overwhelm the burst in low
cutoff data shown in Fig.\,\ref{fig2}b. But for ten NMs with
higher cutoffs (mean R$\rm _c$\,=\,6.4\,GV), the IA amplitude
might not have been too large, allowing the burst to
stay visible as seen from Fig.\,\ref{fig2}a. A power-law
behavior results in a strong dependence of the IA on R$\rm _c$.
But the burst being caused by a change in cutoff rigidity
`$\Delta$R$\rm _c$' shows almost no dependence on R$\rm _c$.
Furthermore, due to high R$\rm _c$ the IA contribution in
GRAPES-3 was small, making it sensitive to burst-like activity.
To summarize, the IA, and burst are two distinct phenomena,
the IA because of its interplanetary origin displays a
direction dependent arrival time, and its amplitude shows a
strong dependence on R$\rm _c$. The burst due to its
local origin close to the Earth occurs simultaneously in
all nine directions, and its amplitude does not depend on
R$\rm _c$. The strong cutoff rigidity dependence of
the IA offers a possible explanation for the inconsistent
observation of the burst by the NMs.

\section{New simulations of the GRAPES-3 burst}
\begin{figure}[]
\begin{center}
\includegraphics*[width=0.49\textwidth,angle=0,clip]{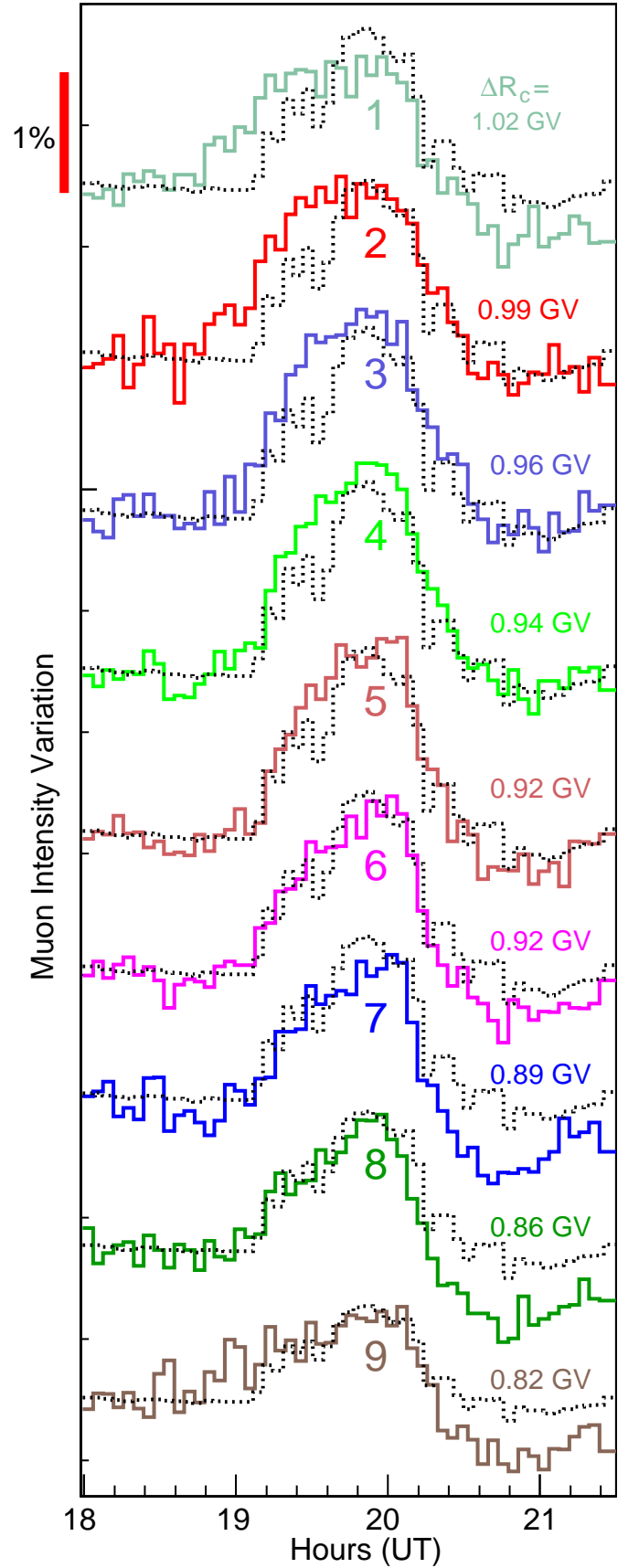}
\vskip -0.05in
\caption{\label{fig4} Muon intensity variation observed by GRAPES-3
                      on 22 June 2015 in nine directions labeled `1'
                      in East through to `9' in West. Monte Carlo
                      simulation for each direction is shown by
                      dotted lines.}
\end{center}
\vskip -0.3in
\end{figure}

\vskip -0.2in
Due to a slowly changing profile, the FD predominantly
contributes to frequencies below 0.5 cycle per day (CPD),
and the burst above 3.5\,CPD. Thus, by designing,
and implementing a suitable fast Fourier transform (FFT)
based filter, the burst can be easily extracted. The
data for 2$^{13}$\,=\,8192 intervals of 4\,min each spanning
23 days (12 June 2015 18:28\,UT -- 4 July 2015 12:36\,UT)
were used, as before \cite{Mohanty16}. The muon data were
analyzed by applying a filter to reject contributions from
the FD, and IA by excluding frequencies below 3.5\,CPD in
the FFT spectrum. The inverse FFT of the filtered spectra
for the nine E-W directions is shown in Fig.\,\ref{fig4}.
The burst is visible in each direction that are labeled 1
through 9. Also shown on each plot, is the result of Monte
Carlo simulations of the burst by superposed dotted lines
as elaborated below.

In our previous work, the occurrence of the burst was
explained by a lowering of the cutoff rigidity
R$\rm _c$ which in turn was caused by the weakening of the
GMF due to its reconnection with the IMF
\cite{Mohanty16}. In a recent monograph, it was highlighted
that typical solar wind speed V$_{\rm SW}$ is $\sim$400\,km/s
when approaching the Earth, faster than typical waves in the
solar wind including the fast magnetosonic speed V$_{\rm MS}$.
At the bow shock, the plasma flow abruptly decreases over a
short distance, with a corresponding increase in the
plasma density, temperature, and magnetic field. For
V$_{\rm SW}$ far above V$_{\rm MS}$, the V$_{\rm SW}$ decreases
by a factor of 4 while the density, and magnetic field
increase by the same factor \cite{Parker16}. In the present
case, V$_{\rm SW}$ being 700\,km/s clearly fulfills this
criterion. New simulations of the burst were carried out by
implementing the above criterion by increasing the magnitude
of the IMF by a factor of 4 for the duration of this
study. The cutoff rigidities for the nine directions as
shown in Fig.\ref{fig1}b were calculated by tracing particle
trajectories \cite{Smart05} in a GMF modeled by IGRF-11
\cite{IGRF11} as detailed in \cite{Mohanty16}. The
time-dependent cutoff rigidities modified by a varying IMF
were calculated every 4\,min after adding 4$\times$IMF
(4B$\rm _x$, 4B$\rm _y$, 4B$\rm _z$) to the respective GMF
components \cite{Mohanty16}. 

The atmospheric muons produced by the GCRs above R$\rm _{c}$
were simulated by using the Monte Carlo code CORSIKA
\cite{CORSIKA}. Muons meeting the telescope trigger requirement
were binned into nine E-W directions shown in Fig.\,\ref{fig1}b.
The difference in the muon intensity before, and after including
the IMF was obtained every 4\,min. The interval 18:00--18:30\,UT
was used as the baseline to estimate the change in muon intensity
for both the data, and simulations. Simulated burst amplitudes
were smaller than the measured ones. When the simulations
were repeated after enhancing the IMF by a factor
2\,$<$\,f\,$<$6 for the duration of the burst, it showed 
that the amplitude of the effect scaled with `f'. A 28\,min
delay of the simulated profiles maximized their
correlation with the muon intensity profiles. A simultaneous
minimization of $\chi^2$ for the nine pairs of observed, and
simulated profiles yielded f\,=\,(5.25$\pm$0.58), implying an
IMF enhancement by a factor of
4$\times$(5.25$\pm$0.58)\,=\,(21$\pm$2.3).

The simulated profiles are shown in Fig.\,\ref{fig4} by dotted
lines. High correlation coefficients (mean\,=\,0.91$\pm$0.03)
between the two profiles are seen in all nine cases. The
maximum reduction in R$\rm _c$ was found to vary from 1.02 in
the East to 0.82\,GV in the West. Thus, a simple model of the
IMF enhancement by a factor of 21$\pm$2.3 (B$\rm _z$\,=\,-840$
\pm$90\,nT) inside the shock front reproduced the amplitude,
and the shape of each profile remarkably well. The burst
amplitude shows a gradual increase from East to West, reaching
a maximum in direction `5', and then decreasing again. This
behavior is also reproduced in the simulated profiles. The
concurrent change in R$\rm _c$ offers a natural explanation
for simultaneous 2\,h GCRI surge in all nine directions. A mean
offset of only (-1.5$\pm$2)\,min among the nine directions
obtained from a cross correlation is consistent with a zero
offset within the 4\,min telescope resolution. For an
interplanetary phenomenon, the expected time offset between
any two directions should increase with the angle between them,
reaching about 5.5\,h for `1,' and `9'. The near simultaneity
of the burst in all nine directions supports its origin close
to the Earth, well within the magnetosphere.

\section{Discussion}
\vskip -0.2in
A new analysis of the GRAPES-3 data by segmenting its FOV into
nine 10.6$^{\circ}$ wide E-W directions showed a clear presence
of an IA along with the burst. The IA displayed a progressive
shift in phase when viewed from directions `1' through `9', as
expected for an interplanetary phenomenon. The strong rigidity
dependence of IA seen by the GRAPES-3 offers a possible
explanation for the detection of burst by ten NMs operating
at high R$\rm _c$ (mean\,=\,6.4\,GV), and not by 21 NMs at low
R$\rm _c$ (mean\,=\,1.8\,GV) \cite{Evenson17}. The effectiveness
of the FFT based filter as a tool to isolate the burst by
removing the IA as seen from Fig.\,\ref{fig4} may be exploited
by others to detect burst-like activity in their data. It is
to be noted that during the storm both the data, and simulations
display nearly identical behavior as seen from Fig.\,\ref{fig4}.
An enhancement of the IMF by a factor of (21$\pm$2.3) is required
for the simulations to reproduce the data. This value is within
1.2$\sigma$ of the old value of 17 reported earlier from a
different analysis scheme \cite{Mohanty16}. The change in the
cutoff rigidity `$\Delta$R$\rm _c$' gradually decreases by about
20\% from 1.02 to 0.82\,GV for directions `1' through `9'. A
correlation coefficient of 0.91 between the nine sets of data,
and simulated profiles implies a significance of nearly
14\,$\sigma$, supporting the hypothesis that the burst was caused
by the lowering of R$\rm _c$ due to the storm.

As noted before the solar wind speed V$_{\rm SW}$ is supersonic
that abruptly slows down by a factor of 4, and the IMF increases
by the same factor at the bow-shock \cite{Parker16,Kulsrud04}.
However, the arrival of the CME shock front on 22 June 2015,
18:40\,UT caused a compression of the bow shock \cite{Mohanty16},
and triggered magnetic reconnection. Generally, the reconnection
front moves slower than the Alfv{\'en} speed \cite{Parker73}.
The CME would have traveled unimpeded from L1 to the bow shock
at $\sim$11\,R$\rm _E$ \cite{Omniweb}. Thus, the 28\,min delay
measured by the GRAPES-3 would have occurred after crossing
the bow-shock. This delay implies an average speed of $\sim$35\,km/s
for the reconnection front. This is considerably sub-Alfv{\'enic},
and much slower than the near-Earth CME speed of 700\,km/s. The
high fidelity reproduction of the burst by its simulated profile
(91\% correlation) supports the hypothesis of IMF enhancement
($\times$4), and compression ($\times$5.25) due to the
interaction of the CME shock with the bow shock. Such bursts offer
a unique phenomenological probe of a highly turbulent environment
created by two interacting shock fronts. It is clearly shown
that in large solar particle events the muon intensity at GRAPES-3,
and the response of the neutron monitors around the world depends
on two distinct phenomena, (i) changes in the local cutoff rigidity
due to reconnection processes, and (ii) anisotropies of
interplanetary cosmic-ray intensities. This recognition should
lead to improved interpretation and coordination of these related
data sets.

It should be noted that the measurement of the CME
properties at L1 by the satellites is an essential
requirement for obtaining an initial estimate of
the storm arrival time on Earth. Since the burst is caused by
the change in GMF in the vicinity of Earth, the start time of
the storm estimated from the burst is likely to be more accurate
than the extrapolation of the CME measurements from L1.
Here it should be emphasized that the present work is the
outcome of a post facto analysis.
The GRAPES-3 telescope has collected uninterrupted data since
early 1999, which is being analyzed to extract more burst-like
events. The discovery of more bursts associated with CMEs
of different speeds, and IMF values should lead to a better
understanding of the storm arrival time on Earth with the
potential for real-time space weather forecasts. Since the only
known preventive measure to avoid damage to modern space, and
ground based technological assets by a super-storm is to disable
their electrical supply, therefore, any development leading to a
better estimate of the arrival time of future super-storms is
highly desirable.

\vskip -0.2in
\section{Conclusions}
\vskip -0.1in
The GRAPES-3 muon telescope in Ooty, India had reported the
detection of a 2\,h burst of 20\,GeV GCRs starting 22 June
2015 19:00\,UT. A new analysis of this data in nine E-W
directions with 10.6$^{\circ}$ segmentation showed the burst
was accompanied by an IA with a strong rigidity dependence
that can naturally explain the inconsistent detection
of the burst by 31 NMs located across the globe. 
Based on
measured 28\,min delay of the storm during the 22 June 2015
burst indicates that the reconnection front in the
magnetosphere of the Earth was moving with a speed of
35\,km/s. The discovery of more burst-like events in
the existing 19 years of data may be helpful in providing a
better estimate of the arrival time of future super-storms.

\section {Acknowledgments} 
\vskip -0.2in
We thank D.B. Arjunan, V. Jeyakumar, S. Kingston, K. Manjunath,
S. Murugapandian, S. Pandurangan, B. Rajesh, K. Ramadass, V.
Santoshkumar, M.S. Shareef, C. Shobana, R. Sureshkumar for their
help in running the experiment. We are grateful to Profs. A.
Kakodkar, K. Kasturirangan, B.V. Sreekantan, and M.R. Srinivasan
for inspiring, and stimulating discussions. 



\begin{thebibliography}{99}

\bibitem{Mohanty16}
P.K. Mohanty et al., Phys. Rev. Lett. {\bf 117}, 171101 (2016).

\bibitem{Evenson17}
P. Evenson et al., Proc. 35th Int. Cosmic Ray. Conf. Busan, PoS(ICRC2017)133 (2017).

\bibitem{Olsen15}
N. Olsen et al., Geophys. Res. Lett. {\bf 42}, 1092 (2015).

\bibitem{Dorman09}
L.I. Dorman, Cosmic Rays in Magnetosphere of Earth and Planets,
Springer, ISBN 978-1-4020-9239-8 (2009).

\bibitem{Webb12}
D.F. Webb, T.A. Howard Living Rev. Solar Phys. {\bf 9}, 3 (2012).

\bibitem{Kudela00}
K. Kudela, M. Storini, M.Y. Hofer, and A. Belov, Rev. Space Sci. {\bf 93}, 153 (2000).

\bibitem{Carrington59}
R.C. Carrington, Mon. Not. R. Astron. Soc. {\bf 20}, 13 (1859).

\bibitem{Liu14}
Y.D. Liu et al., Nature Comm. {\bf 5}, 3481 (2014).

\bibitem{NASA08}
Severe Space Weather Events--Understanding Societal and Economic Impacts: A Workshop Report, DOI 10.17226/12507

\bibitem{OECD11}
http://www.oecd.org/gov/risk/46891645.pdf

\bibitem{Forbush38}
S. E. Forbush, Phys. Rev. 54, 975 (1938);
H.V. Cane, Space Sci. Rev. {\bf 93}, 55 (2000);
P. Subramanian et al., Astron. Astrophys. {\bf 494}, 1107 (2009);
K.P. Arunbabu et al., Astron.  Astrophys. {\bf 555}, A139 (2013);
K.P. Arunbabu et al., Astron.  Astrophys. {\bf 580}, A41 (2015).

\bibitem{Smart05}
D.F. Smart and M.A. Shea, Adv. Space Res. {\bf 36}, 2012 (2005).

\bibitem{Kondo62}
I. Kondo, J. Phys. Soc. Japan {\bf 17}, 402 (1962).

\bibitem{Kudo87}
S. Kudo et al., J. Geophys. Res. {\bf 92}, 4719 (1987).

\bibitem{Hayashi05}
Y. Hayashi et al., Nucl. Instrum. Meth. A {\bf 545}, 643 (2005).

\bibitem{Nonaka06}
T. Nonaka et al., Phys. Rev. D {\bf 74}, 052003 (2006).

\bibitem{Omniweb}
http://omniweb.gsfc.nasa.gov/form/omni\_min.html

\bibitem{Bargatze05}
L.F. Bargatze et al., J. Geophys. Res. {\bf 110}, A07105, (2005).

\bibitem{Mohanty15}
P.K. Mohanty et al., Proc. of Science (ICRC2015) 045;
P.K. Mohanty et al., Astropart. Phys. {\bf 79} 23 (2016).

\bibitem{Parker16}
Magnetic Reconnection, Concepts and Applications, Springer,
ISBN: 978-3-319-26430-1 Ed: W.D. Gonzalez, and E.N. Parker, (2016).

\bibitem{IGRF11}
C. Finlay et al., Geophys. J. Int. {\bf 183}, 1216 (2010).

\bibitem{CORSIKA}
https://www.ikp.kit.edu/corsika/

\bibitem{Kulsrud04}
R.M. Kulsrud, Plasma Physics for Astrophysics, Princeton U. Press,
ISBN: 978-0691120737 (2004).

\bibitem{Parker73}
E.N. Parker, Astrophys. J. {\bf 180}, 247 (1973);
K.P. Dere, Astrophys. J. {\bf 472}, 864 (1996);
L. Comisso, and A. Bhattacharjee, J. Plasma Phys. {\bf 82}, 595820601 (2016);
P.A. Cassak, Y.-H. Liu and M.A. Shay, J. Plasma Phys. {\bf 83}, 715830501 (2017).
\end{thebibliography}
\end{document}